\documentclass[10pt,prx,twocolumn,amsmath,amssymb,floatfix,notitlepage,superscriptaddress,groupedaddress]{revtex4-2}
\usepackage[T1]{fontenc}
\usepackage[utf8]{inputenc}
\usepackage{lmodern}
\usepackage{microtype,bm,bbm}
\usepackage{graphicx,booktabs}
\usepackage{times}
\usepackage[usenames,dvipsnames]{xcolor}
\usepackage[english]{babel}
\usepackage{upgreek}
\usepackage{braket}
\usepackage{subfigure}
\usepackage{dsfont}
\usepackage{transparent}
\usepackage{layouts}
\usepackage{tikz}
\usepackage{adjustbox}

\usepackage{printlen}

\usepackage{hyperref}
\hypersetup{
  colorlinks,
  allcolors=NavyBlue,
  linktoc=all
}

\usepackage[capitalize]{cleveref}
\usepackage{pifont}
\crefformat{section}{Sec.~#2#1#3}

\graphicspath{ {figures/}}
\graphicspath{ {inkscape_figures/}}

\begin{document}

\title{Quench Spectroscopy for Two-Dimensional Spin Models with Power-Law Interactions}

\author{Katharina Brechtelsbauer}
\author{Hans Peter Büchler}
\affiliation{Institute for Theoretical Physics III and Center for Integrated Quantum Science and Technology, University of Stuttgart, Pfaffenwaldring 57, 70569 Stuttgart, Germany}

\date{\today}
\pacs{}

\begin{abstract}

In this paper, we analytically compute the quench dynamics of simple product states that evolve under a spin-1/2 XY-model with power-law interactions in a staggered magnetic field. We use linear spin-wave theory as well as a low-energy theory that accounts for strong phase fluctuations. This allows us to provide an analytic expression for the dynamics of phase-type correlations with short-range interactions, where the linear spin-wave theory is known to fail. For density-type correlations the low-energy theory and the linear spin-wave theory give the same result. Our analysis is valid for arbitrary magnetic field strength and in particular provides an analytical understanding of the quench dynamics near criticality. Our results can thus provide useful tools for the experimental observation of excitation gaps and critical behavior.

\end{abstract}
\maketitle

\section{Introduction}

The experimental progress in the development of quantum simulation platforms such as Rydberg atoms, trapped ions, polar molecules, highly magnetic atoms or atoms coupled to cavities \cite{Browaeys2020,Monroe2021,Bao2023,Chomaz2023,Ritsch2013} brought attention to systems with interactions extending beyond short range interactions \cite{Defenu2021}. Such long-range  interactions can strongly affect the physical properties of the system and can give rise to unique phenomena; for example Wigner crystals that form in the presence of Coulomb interactions \cite{Bonsall1977}.  One distinguishes three different regimes to describe the character of the interaction: the \textit{strong long-range} regime, where the energy of the system is superextensive, the \textit{weak long-range} regime, where the energy is extensive, but the characteristic properties are affected by the interaction range, and the \textit{short-range regime}, where the system shows the same behavior  as systems with short-range interactions \cite{Defenu2021}. The observation of these phenomena requires among others the measurement of excitation spectra, and recently a  promising approach emerged to measure excitation spectra  using quenches \cite{Calabrese2006,Gritsev2007,Hauke2013,Cevolani2015,Frrot2018,Menu2018,Cevolani2018,Schemmer2018,Villa2019,Villa2020,Menu2023,Bocini2025}. Here, we consider the quench dynamics of a ferromagnetic spin-1/2 XY-model with algebraic interactions.

The ferromagnetic XY-model with its exchange interactions is well-suited for studying the different interaction regimes. In particular, the two-dimensional, ferromagnetic XY-model with dipolar interactions is in the weak long-range regime. In this case the spin-wave dispersion scales as $\omega_\mathbf{k}\propto \sqrt{|\mathbf{k}|}$ for small wave vectors $\mathbf{k}$, while for the short-range model one finds a linear dispersion relation \cite{Peter2012}. 
This weak-long range interaction alters the upper critical dimension and especially the two-dimensional ferromagnetic XY-model is well described by mean-field theory \cite{Fisher1972,Sak1973,Defenu2020,Defenu2021}
with strongly modified critical exponents at the phase transition. Remarkably, even superfluid transport properties of the system are affected \cite{Yabuuchi2025}.
While in low-dimensional systems with local interactions fluctuations prevent ordering \cite{Mermin1966}, dipolar interactions can stabilize the formation of order in two-dimensions \cite{Bruno2001}. Recently, the presence of long-range order and the square-root behavior of the excitations in the ferromagnetic XY-model have been observed in an array of Rydberg atoms \cite{Chen2023,Chen2025}. The measurement of the excitation spectrum utilized a method called quench spectroscopy \cite{Chen2025}. Here, the system is initially not prepared in an eigenstate and the excitation spectrum is extracted from the propagation of quasiparticles \cite{Calabrese2006}. In the weak long-range interaction regime, the dynamics can be derived with linear spin-wave theory, while for short-range interactions the derivation is not self-consistent due to the appearance of infrared divergencies \cite{Frrot2018}. Nevertheless, in reference \cite{Chen2025} the method was successfully used to experimentally probe the excitation spectrum of the antiferromagnetic XY-model with dipolar interactions, which serves as a representative for the short-range regime.

\begin{figure}[tb]
  \includegraphics[width=\linewidth]{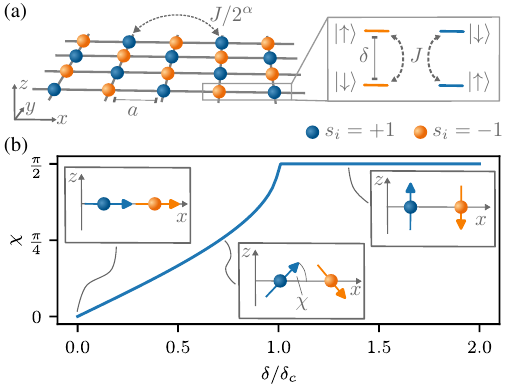}
  \caption{Illustration of the model (panel (a)) and the spin expectation values of the variational mean-field states (panel (b)). In this paper we derive how the mean-field states in (b) evolve under the model in (a). (a) We consider spin-1/2 particles on a 2-dimensional square lattice with lattice constant $a$, that interact via power-law exchange interaction. The spins are exposed to a staggered magnetic field with strength $\delta$ perpendicular to the lattice. The different colors label the sign of the magnetic field. (b) For the quench dynamics we analyze the time-evolution of mean-field states. These states are simple product states that minimize the energy expectation value. As variational parameter we choose the angle $\chi$ between the spin-expectation value and the x-axis.  }
  \label{fig:illustrationmeanfield}
\end{figure}

Here, we use a low-energy effective theory to account for the strong phase fluctuations that lead to divergences in the linear spin-wave analysis with short-range interactions. In addition, we extend the system by a staggered magnetic field - as illustrated in \cref{fig:illustrationmeanfield}a - and generalize both, the linear spin-wave derivation of reference \cite{Frrot2018} and the low-energy theory to non-zero magnetic fields.
Our results are two-fold. On the one hand we find that the dynamics of density-type correlations at any time is well described by linear spin-wave theory even for short-range interactions, despite the fact that the linearization is in principle not valid in this case. And furthermore, from the low-energy theory we obtain an analytic expression for the quench dynamics of phase-type correlations in the short-range regime.  On the other hand, our analysis gives an analytical understanding of the quench dynamics of the system for arbitrary magnetic fields and thus allows the observation of excitation gaps near the quantum phase transition from the ferromagnetic phase into the paramagnetic phase.

The rest of this work is structured as follows. In \cref{sec:setup} we introduce the model, the phase-diagram and the variational mean-field state. In \cref{sec:lsw} we use linear spin-wave theory to compute the quench dynamics of the system initially prepared in the mean-field state and extend the results of reference \cite{Frrot2018} to arbitrary magnetic field strength. Since the linear spin-wave result for the quench dynamics is not self-consistent for short-range interactions, we introduce a low-energy theory in \cref{sec:lowenergy} and use this to again compute the quench dynamics. We conclude in \cref{sec:conclusion}.

\section{Setup}
\label{sec:setup}
Our goal is to derive the dynamics of a spin-1/2 system initially prepared in a simple product state. 
In this section, we introduce the model and the initial state.

\subsection{Model and Phase Diagram} 
Inspired by recent experiments \cite{Chen2023}, we consider a two-dimensional spin-1/2 XY-model with power-law interactions in a staggered magnetic field as illustrated in \cref{fig:illustrationmeanfield}a. The Hamiltonian of this system is given by 
\begin{align}
H = -  J\sum_{i\neq j} \frac{a^\alpha}{|\mathbf{r}_i-\mathbf{r}_j|^\alpha} \left( S^x_i S^x_j +S^y_iS^y_j \right) -  \delta \sum_i s_i S^z_i,
\label{eq:hamiltonian}
\end{align}
where $S^x_i$, $S^y_i$, and $S^z_i$ are the components of the spin-operator acting on the $i$-th spin, $J>0$ is the interaction strength, $a$ the lattice constant, $\delta$ is the strength of  the staggered magnetic field and $s_i = \pm 1$, depending on the sublattice of the lattice site $i$. The case with $\alpha=3$ corresponds to dipolar exchange interactions that are, for example, present in  experiments with Rydberg atoms or polar molecules \cite{Browaeys2020,Bao2023}. In the limit $\alpha \rightarrow \infty$ one obtains nearest-neighbor interactions.

First, we summarize the zero-temperature ground state phase diagram \cite{Chen2023,Sbierski2024}. The system exhibits a phase transition from a gapped paramagnetic phase for large $\delta/J$ to an XY-ferromagnetic phase with long-range order and gapless excitations (small $\delta /J$). The long-range order can be observed from non-vanishing correlations $\langle S^x_i S^x_j +S^y_i S^y_j\rangle$. The excitation spectrum in the XY-ferromagnetic phase is strongly affected by the interaction range. For $\alpha>4$ one finds a linear dispersion for small $k$, while for $2<\alpha<4$ the low-energy dispersion goes with $k^{(\alpha-2)/2}$ \cite{Peter2012,Song2023,Diessel2023}. For $\alpha=4$ the dispersion is linear with logarithmic corrections \cite{Diessel2023}. 

The $\alpha$-dependence of the Goldstone mode correlates with the critical behavior of the system and furthermore with the system properties at finite-temperature. For $2<\alpha<4$, the critical behavior is well described by mean-field theory, while for $\alpha>4$ the critical exponents differ from the mean-field values \cite{Defenu2021}. At finite temperature the linear dispersion gives rise to logarithmic divergences in the Green's function and allows for strong fluctuations that prevent the formation of order \cite{Mermin1966,Bruno2001}. For $2<\alpha<4$ no divergences appear and one still can observe XY-ferromagnetic ordering \cite{Bruno2001,Chen2023}. Due to such logarithmic divergences, also the linear spin-wave description of the quench dynamics is not self-consistent \cite{Frrot2018}, as we discuss later in more detail. For any $\alpha>4$ the leading contributions in the effective field theory are the same as for nearest neighbor interactions and the system is short-range. For $2<\alpha<4$ the interactions are weak long-range and for $\alpha<2$ one finds strong long-range interactions \cite{Fisher1972,Sak1973,Defenu2020,Defenu2021}.

\subsection{Variational Mean-Field State} 
For $\delta=0$ the mean-field ground state is given by $\ket{\psi_0}=\ket{\rightarrow}^{\otimes N}$, where all spins are aligned along the x-direction \cite{Frrot2018}. When we switch on the magnetic field, the spins also start to pick up an alternating alignment along 
the z-direction. We account for this by rotating the spin-operators around the y-axis by an angle $\pm \chi$, such that the mean-field state for non-zero $\delta$ is $\ket{\psi}=\prod_j e^{i\chi s_j S^y_j} \ket{\psi_0}$.
The variational parameter $\chi$ is obtained by minimizing the energy of the variational mean-field wave function
\begin{align}
\bra{\psi} H \ket{\psi}= \frac{N}{4} \left( -J \gamma_0 \cos^2(\chi) - 2 \delta\sin(\chi) \right).
\end{align}
Here, $\gamma_0$ is the $\mathbf{k}=0$  component of the Fourier transform of the interaction $\gamma_\mathbf{k}=\sum_{\mathbf{r}\neq 0} (a/|\mathbf{r}|)^{\alpha} e^{i \mathbf{k} \cdot \mathbf{r}}$. 
For $\delta>\delta_c:=J\gamma_0$ one finds $\chi=\pi/2$, which gives the paramagnetic state, where spins align along the staggered magnetic field. For $\delta<\delta_c$ one finds $\chi=\arcsin(\delta/\delta_c)$. The spin expectation values of the mean-field ground states $\ket{\psi}$ are illustrated in \cref{fig:illustrationmeanfield}b.

In the following we analyze the quench dynamics for the system initially prepared in this mean-field state. 
First, we revisit the linear spin-wave results from reference \cite{Frrot2018} and extend them to arbitrary $\delta$. As we will see, the linear spin-wave analysis of the quench dynamics is not self-consistent for short-range interactions \cite{Frrot2018}. Thus, we additionally introduce a low-energy effective theory that allows us to derive analytical expressions for the long-time dynamics for  arbitrary $\alpha$.

\section{Linear spin-wave theory}  
\label{sec:lsw}

In this section, we extend the results of reference \cite{Frrot2018} to non-zero magnetic fields $\delta$ and show that for short-range interactions the derivation of the quench dynamics is not self-consistent in the ordered phase and close to the phase transition. 

\subsection{Excitation Spectrum}
To derive the excitation spectrum and the quench dynamics, we use the Holstein-Primakoff transformation to express the spin operators in terms of bosonic creation and annihilation operators $b_i^\dagger$ and $b_i$. For $\delta=0$, the spin operators then read

\begin{align}
S_{i,0}^x&=\frac{1}{2}-b_i^\dagger b_i,    \\
S_{i,0}^y&=\frac{1}{2 i} \left(\sqrt{1\!-\!b_i^\dagger b_i} \:  b_i - b_i^{\dag }\sqrt{1\!-\!b_i^\dagger b_i} \right)\approx \frac{b_i-b_i^\dagger}{2 i},\ \notag \\
S_{i,0}^z&=-\frac{1}{2 } \left(\sqrt{1\!-\!b_i^\dagger b_i}\: b_i + b_i^{\dag}\sqrt{1\!-\!b_i^\dagger b_i} \right) \approx -\frac{b_i+b_i^\dagger}{2 } \notag.
\end{align}
For arbitrary $\delta$  this generalizes to $\mathbf{S}_{i}=R_y(\pm \chi) \mathbf{S}_{i,0}$, where we rotate $\mathbf{S}_{i,0}$ by the angle $\chi$ around the y-axis, such that the mean-field state $\ket{\psi}$ contains no Holstein-Primakoff bosons. 

To account for the two sublattices, that have to be distinguished due to the staggered magnetic field, we introduce two types of Holstein-Primakoff operators. The operators $a_i$ and $a_i^\dagger$ act on one sublattice, while the operators $b_i$ and $b_i^\dagger$ act on the other one. In leading order the effective Hamiltonian then takes the form 
\begin{align}
H_\text{eff}=\frac{1}{2} \sum_\mathbf{k} (\mathbf{a}_{\mathbf{k}}^\dagger,\mathbf{b}_{\mathbf{k}}^\dagger) \begin{pmatrix} M_1 & M_2 \\ M_2 & M_1 \end{pmatrix} \begin{pmatrix} \mathbf{a}_{\mathbf{k}} \\ \mathbf{b}_{\mathbf{k}} \end{pmatrix}
\label{eq:lswhamiltonian}
\end{align}
with $\mathbf{a}_{k}^\dagger=(a_\mathbf{k}^\dagger,a_{-\mathbf{k}})$ and $a_\mathbf{k}$ the Fourier transform of $a_i$. The operator $\mathbf{b}_{\mathbf{k}}^\dagger$ is defined analogously and $M_1$ and $M_2$ are given by
\begin{align} 
M_1&=\gamma_{1,\mathbf{k}}\begin{pmatrix} d & g \\ g & d \end{pmatrix}+c_0 \mathbbm{1}, \notag \\
M_2&=-\gamma_{2,\mathbf{k}}\begin{pmatrix} g & d \\ d & g \end{pmatrix}.
\end{align}
Here, $\gamma_{1,\mathbf{k}}$ and $\gamma_{2,\mathbf{k}}$ are the Fourier transform of the interactions within each sublattice with $\gamma_\mathbf{k}=\gamma_{1,\mathbf{k}}+\gamma_{2,\mathbf{k}}$, and $g=J/2 \cos^2(\chi)$, $d=-J/2(1+\sin^2(\chi))$ and $c_0=\delta \sin(\chi)+J \gamma_0 \cos^2(\chi)$. 

The Hamiltonian can be diagonalized by first transforming to the symmetric and antisymmetric combinations  $a_{\mathbf{k}}\pm b_{\mathbf{k}}$, and afterward applying the Bogoliubov transformation. The spectrum is then given by $\hbar \omega_{\pm,\mathbf{k}}=\sqrt{A_{\pm,\mathbf{k}}^2-B_{\pm,\mathbf{k}}^2}$ with $A_{\pm,\mathbf{k}}=c_0+d\gamma_{1,\mathbf{k}}\mp g\gamma_{2,\mathbf{k}}$ and $B_{\pm,\mathbf{k}}=g\gamma_{1,\mathbf{k}}\mp d \gamma_{2,\mathbf{k}}$.
In the XY-ferromagnetic phase one finds one gapped energy band and a gapless Goldstone mode with low-energy dispersion $\omega_{+,\mathbf{\mathbf{k}}}=(v |\mathbf{k}|)^{\sigma/2}$ with $\sigma=2$ for $\alpha> 4$ and $\sigma=\alpha-2$ for $2<\alpha<4$, as expected. For $\alpha=4$ the low-energy dispersion is linear with logarithmic corrections. In the paramagnetic phase the excitations are gapped and degenerate. 

\subsection{Quench Dynamics}

Finally, we  use these results to derive an analytic expression for the quench dynamics for arbitrary $\delta$. The quantities we are interested in are $\mathcal{S}_{ij}^{\mu\mu}(t)= \bra{\psi} S^\mu_i(t) S^\mu_{j}(t) \ket{\psi}$ with \mbox{$\mu\in \{y,\perp\}$}. Here, $S^\perp= \sin(\chi) S^x \mp \cos(\chi) S^z$  is the component of the spin operator pointing orthogonal to the y-direction and the direction of the mean-field spin expectation value. For $\delta=0$ this is simply $\mp S^z$.

To account for the different sublattices it is useful to introduce the Fourier transformation within one sublattice as $\mathcal{S}_{\mathbf{k},AA}^{yy}(t)= \frac{2}{N} \sum_{i,j \in A} e^{i \mathbf{k} \cdot \mathbf{r}_{ij}} \mathcal{S}_{ij}^{yy}(t)$ and between two sublattices $\mathcal{S}_{\mathbf{k},AB}^{yy}(t)= \frac{2}{N} \sum_{i \in A,j \in B} e^{i \mathbf{k} \cdot \mathbf{r}_{ij}} \mathcal{S}_{ij}^{yy}(t)$. Note that due to symmetry $\mathcal{S}_{\mathbf{k},AA}(t)=\mathcal{S}_{\mathbf{k},BB}(t)$. This allows us to define $\mathcal{S}_{\pm,\mathbf{k}}^{yy}(t)=(\mathcal{S}_{\mathbf{k},AA}^{yy}(t) \pm \mathcal{S}_{\mathbf{k},AB}^{yy}(t))$.
Note that $\mathcal{S}_{+,\mathbf{k}}^{yy}(t)$ corresponds to the definition of the structure factor in reference \cite{Frrot2018}.

The quench dynamics of the structure factor is then calculated analogously to reference \cite{Frrot2018} and one finds 
\begin{align}
\mathcal{S}_{\pm,\mathbf{k}}^{yy}(t)=\frac{1}{4} \left(1+ 2 \frac{B_{\pm,\mathbf{k}}^2+A_{\pm,\mathbf{k}}B_{\pm,\mathbf{k}}}{\omega_{\pm,\mathbf{k}}^2} \sin^2(\omega_{\pm,\mathbf{k}} t)\right), \notag \\
\mathcal{S}_{\pm,\mathbf{k}}^{\perp\perp}(t)=\frac{1}{4} \left(1+ 2 \frac{B_{\pm,\mathbf{k}}^2-A_{\pm,\mathbf{k}}B_{\pm,
\mathbf{k}}}{\omega_{\pm,\mathbf{k}}^2} \sin^2(\omega_{\pm,\mathbf{k}} t)\right).
\label{eq:corrresultperplsw}
\end{align}

\subsection{Validity of the Approximations}

\begin{figure}[tb]
  \includegraphics[width=\linewidth]{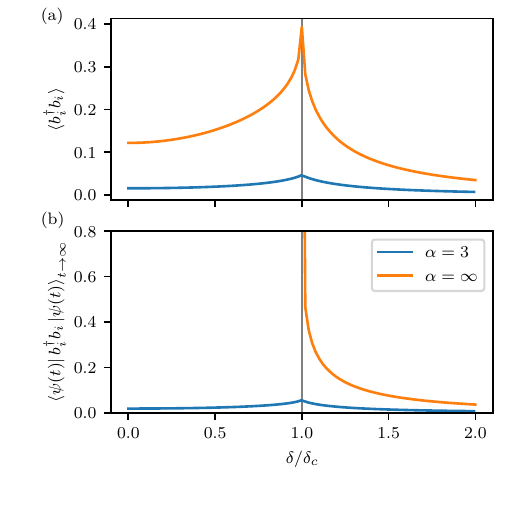}
  \caption{Numerical results for the expressions in \cref{eq:validitylswgsfluctutations} and \cref{eq:validitylswquench}. Panel (a) shows the number of Holstein-Primakoff (HP) bosons in the ground state and panel (b) the number of HP-bosons in the time evolved initial state in the limit for $t\rightarrow \infty$ for different magnetic field strengths. The case with $\alpha=3$ represents the weak long-range regime, while $\alpha=\infty$ corresponds to short-range interactions. }
  \label{fig:validitylsw}
\end{figure}

In the previous derivation we made use of two approximations. First, we assumed that fluctuations on top of the ground state are small, that is  $\langle b_i^\dagger b_i \rangle \ll 1$. This allowed us to derive the quadratic form of the Hamiltonian as given in \cref{eq:lswhamiltonian}. Second, when deriving the quench dynamics of the spin-spin correlations, we again linearized the Holstein-Primakoff transformation. This is only valid if $\bra{\psi(t)} b_i^\dagger b_i \ket{\psi(t)} \ll 1$. As pointed out in reference \cite{Frrot2018} for $\delta=0$, the second approximation breaks down for $\alpha>4$. In the following, we analyze this in more detail. 

Within linear spin-wave theory one obtains 
\begin{align}
 \langle b_i^\dagger b_i \rangle &=  \frac{1}{2N_A}\sum_{\mathbf{k},\sigma} \left(\frac{A_{\sigma,\mathbf{k}}}{\omega_{\sigma,\mathbf{k}}} -1 \right) 
 \label{eq:validitylswgsfluctutations}
\end{align}
and
\begin{align}
 \bra{\psi(t)} b_i^\dagger b_i \ket{\psi(t)} =\frac{1}{2 N_A}\sum_{\mathbf{k},\sigma}\frac{B_{\sigma,\mathbf{k}}^2}{\omega_{\sigma,\mathbf{k}}^2} \sin^2(\omega_{\sigma,\mathbf{k}}t),
  \label{eq:validitylswquench}
\end{align}
where $N_A=N/2$ is the number of lattice sites in one sublattice.
When transferring the sum into a 2D integral, we immediately see that in the second expression infrared divergences appear in the short-range case with $\omega_{+,\mathbf{k}}\propto k$, while in the first expression the integrand is still finite for $k\rightarrow 0$. Furthermore, also in the gapped phase near the phase transition, the second expression diverges logarithmically for short-range interactions. The reason is that the mean-field state $\ket{\psi}$ has admixed some excited states, which gives rise to strong phase-fluctuations. 

In \cref{fig:validitylsw} we exemplarily compute both expectation values numerically for $\alpha=3$ and $\alpha=\infty$ and $t\rightarrow \infty$, where $\sin^2(\omega_{\sigma,\mathbf{k}} t)$ is replaced with $1/2$. As expected, the fluctuations during the quench dynamics diverge in the ordered phase and close to the phase transition. The ground state fluctuations remain finite for all interaction ranges and reach their maximum at $\delta=\delta_c$. However, for short-range interactions the fluctuations are stronger, which agrees well with the fact that for weak long range interactions the system is expected to behave mean-field like, while for short-range interactions the critical exponents differ from mean-field theory.

In summary, the linear spin-wave analysis of the quench dynamics is not self-consistent in the case of short-range interactions. To deal with this, we compute the quench dynamics within a low-energy theory, where we express the spin operators in terms of phase and density fluctuations.

\section{Low-energy effective theory} 
\label{sec:lowenergy}

In this section we derive the quench dynamics within a low-energy effective theory. We first derive an effective Hamiltonian for phase- and density-fluctuations. Afterward, we use this Hamiltonian to derive the dynamics of coarse-grained spin-spin correlations for low energies and compare the result with the linear spin-wave approach. Finally, we calculate the space-time dependence of phase-type correlations with short-range interactions.

\subsection{Effective Hamiltonian}

\begin{figure}[tb]
  \includegraphics[width=\linewidth]{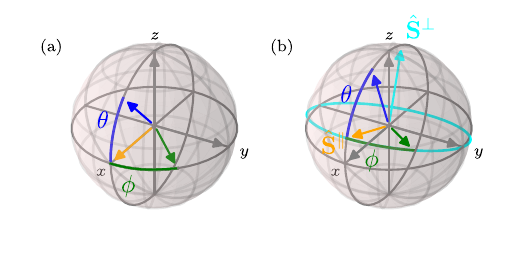}
  \caption{Visualization of the spin expectation values of the spin coherent states $\ket{\theta_i,\phi_i}$ for (a) $\delta=0$ and (b) $0<\delta<\delta_c$. The orange arrow corresponds to the mean-field state $\ket{\psi_i}=\ket{\phi_i=0,\theta_i=0}$. The green and blue arrows illustrate spin expectation values for $\phi\neq0$ and $\theta \neq 0$ respectively. The cyan arrow labels the direction perpendicular to the mean-field spin and the y-direction. $\phi$ describes deflections of the mean-field spin in the plane perpendicular to this vector.}
  \label{fig:loweenergyvisualization}
\end{figure}

To express the spins in terms of phase and density fluctuations we introduce the spin coherent states $\ket{\theta_i,\phi_i}=e^{-i \phi_i S^\perp_i} e^{\pm i \theta_i S^y_i} \ket{\psi_i}$, where $\ket{\psi_i}$ is the mean-field state at the lattice site $i$. The spin expectation value of this state is illustrated in \cref{fig:loweenergyvisualization}. For $\delta=0$ the angles $\theta$ and $\phi$ describe deflections of the mean-field spin in the $x$-$z$-plane and the $x$-$y$-plane. Due to the $U(1)$-symmetry of the XY-model it is thus obvious that $\phi$ gives rise to phase-fluctuations, while $\theta$ can be identified with density fluctuations. This generalizes to nonzero $\delta<\delta_c$, as we will see from the effective Hamiltonian. For $\delta>\delta_c$ the fluctuations in $\phi$ also become density-like.

Using the coherent state path integral formalism \cite{Kuratsuji1992} we derive the effective Hamiltonian for density and phase fluctuations. The Berry phase term allows us to identify  $\sin(\theta)/2$ as the canonical conjugate of $\phi$. This motivates the definition of the density fluctuations as $ \delta n_i= \sin(\theta_i)/2$, such that $[\delta n_{j},\phi_{j'}]=i\delta_{j,j'}$. In leading order in $\phi$ and $\delta n$ the low-energy Hamiltonian takes the form
\begin{align}
\tilde{H} = \frac{1}{2} \sum_\mathbf{k} V_\mathbf{k} |\phi_\mathbf{k}|^2 + U_\mathbf{k} |\delta n_\mathbf{k}|^2.
\label{eq:lowenergyhamiltonian}
\end{align}
Here, $\delta n_\mathbf{k}=1/\sqrt{N}\sum_j \delta n_j \exp(i \mathbf{k} \cdot \mathbf{r}_j)$ and $\phi_\mathbf{k}$ are the Fourier modes of the density and phase fluctuations and the coefficients are $V_{\mathbf{k}}=\frac{1}{2}(J \gamma_0 \cos^2(\chi) +\delta \sin(\chi)-J \gamma_{\mathbf{k}})$ and $U_{\mathbf{k}}=2 (J \gamma_0 \cos^2(\chi) +\delta \sin(\chi)-J \sin^2(\chi)\gamma_{\mathbf{k},-})$, where $\gamma_{\mathbf{k},-}=\gamma_{1,\mathbf{k}}-\gamma_{2,\mathbf{k}}$.  
The excitation spectrum for the harmonic oscillator in \cref{eq:lowenergyhamiltonian} is given by $\hbar \omega_\mathbf{k} =  \sqrt{U_\mathbf{k} V_\mathbf{k}}$. Inserting $U_\mathbf{k} $ and $V_\mathbf{k} $ we thus again find gapless modes for $\delta<\delta_c$ and gapped modes for $\delta>\delta_c$.  

Since there are two spins per unit cell we in principle obtain again two bands. However, since we are interested only in the long-wavelength limit, we neglect the antisymmetric modes and drop the label for the symmetric modes. Nevertheless, it is important to note that both modes match exactly the excitation energies one obtains within linear spin-wave theory even for short-range interactions. This is not a surprise, since we -- as discussed before -- do not expect the spin-wave description to fail at all, but only when deriving the quench dynamics.

\subsection{Quench Dynamics}

 Again, we are interested in equal-time correlations $\bra{\psi} S_i^\mu(t)S_j^\mu(t) \ket{\psi}$. Using the coherent spin state formalism, we find that the correlations in terms of density- and phase-fluctuations read 
\begin{align}
\mathcal{S}^{yy}(\mathbf{r}_{ij},t) &\approx \frac{1}{4} \bra{\psi} \sin\left(\phi_i(t)\right) \sin\left(\phi_j(t)\right) \ket{\psi}, \notag \\
\mathcal{S}^{\perp \perp}(\mathbf{r}_{ij},t) &=  \bra{\psi}\delta n_i(t)\delta n_j(t) \ket{\psi}.
\label{eq:correlationswithphaseanddensity}
\end{align} 

The time evolution of the operators $\phi_i(t)$ and $\delta n_i(t)$ is determined by the Hamiltonian in \cref{eq:lowenergyhamiltonian}. However, this Hamiltonian is only valid for low energies and small $k$ and thus the operators $\phi_i$ and $\delta n_i$ cannot resolve single lattice sites. To account for this we introduce coarse graining and group the $N_0$ spins on one coarse-grained lattice site into one smeared out spin, such that $\phi$ and $\delta n$ are averaged over all spins on that coarse-grained lattice site. The coarse-grained mean-field state is $\ket{\psi}=\tilde{\prod_l} \ket{\psi}_l$, where $\prod_l$ is a product over all coarse-grained lattice sites. The mean-field state of one coarse-grained lattice site can be written as
\begin{align}
\ket{\psi}_l = \frac{1}{2^{N_0/2}}\sum_{m=-N_0/2}^{N_0/2}  \sqrt{\binom{N_0}{m+N_0/2}}\ket{m}_l,
\label{eq:mfcoarsegrained}
\end{align}
where $\ket{m}_l$ is a normalized superposition of all eigenstates of the total magnetization $S_{l,\text{tot}}^\perp$ with eigenvalue $m$. 
For large $N_0$ we can approximate the binomial with a Gaussian and the coarse-grained mean-field state reads \mbox{$\ket{\psi}_l\approx \sum_{m \in \mathbb{Z}} \left(\frac{2}{\pi N_0} \right)^{1/4}e^{-\frac{1}{N_0}m^2}\ket{m}_l$}. Here, we made use of the exponential suppression of terms with $|m|>N_0/2$ and extended the sum in \cref{eq:mfcoarsegrained} to $m\in \mathbb{Z}$.

The time evolved operator $\phi_i(t)$ takes the form $\phi_i(t)=N_0 \sum_l C(\mathbf{r}_{il},t) \phi_l - D(\mathbf{r}_{il},t) \delta n_l$, where $\sum_l$ sums over the coarse-grained lattice. The coefficients are $C(\mathbf{r},t) =  \frac{1}{N}\sum_\mathbf{k}e^{i \mathbf{k}\cdot \mathbf{r}} \cos(\omega_{\mathbf{k}} t)$ and $D(\mathbf{r},t)= \frac{1}{N} \sum_\mathbf{k} e^{i \mathbf{k}\cdot \mathbf{r}} \sqrt{\frac{U_{\mathbf{k}}}{V_{\mathbf{k}}}}\sin(\omega_{\mathbf{k}}  t)$, where $\sum_\mathbf{k}$ is restricted to the coarse-grained Brillouin zone.  

The derivation of $\mathcal{S}^{yy}(\mathbf{r}_{ij},t)$ thus reduces to computing $\bra{\psi}_l e^{i N_0 (C\phi_l-D \delta n_l)} \ket{\psi}_l$ for each coarse-grained lattice site $l$ individually. Here $C$ and $D$ are placeholders for $C(\mathbf{r}_{il},t)\pm C(\mathbf{r}_{jl},t)$ and $D(\mathbf{r}_{il},t)\pm D(\mathbf{r}_{jl},t)$. For large $N_0$ one finds 
\begin{align}
\bra{\psi}_l e^{i N_0 (C\phi_l-D\delta n_l)}\ket{\psi}_l=  e^{-\frac{N_0}{8}(4C^2+D^2)}.
\label{eq:mainresult}
\end{align}
Details on the derivation are given in \cref{app:calcmaincorr}. The quench dynamics of the correlations is then given by
\begin{align}
\mathcal{S}^{yy}(\mathbf{r},t)=\frac{e^{-\frac{1}{4}g_{y,1}(\mathbf{r},t)}}{8} \left(1 -e^{-\frac{1}{2}g_{y,2}(\mathbf{r},t)}   \right)
\label{eq:correlationsb}
\end{align}
with 
\begin{align}
g_{y,1}(\mathbf{r},t) &=\frac{1}{N}\sum_{\mathbf{k}}   \left[4 +\left(\frac{U_{\mathbf{k}}^2}{\omega_\mathbf{k}^2}-4\right) \sin^2(\omega_\mathbf{k} t) \right]\left(1-e^{i\mathbf{k}\cdot\mathbf{r}}\right), \notag \\ 
g_{y,2}(\mathbf{r},t) &=\frac{1}{N}\sum_{\mathbf{k}} \left[4 +\left(\frac{U_{\mathbf{k}}^2}{\omega_\mathbf{k}^2}-4\right) \sin^2(\omega_\mathbf{k} t) \right]e^{i\mathbf{k}\cdot\mathbf{r}}, 
\label{eq:correlationsintegral}
\end{align}
for $r\gg\sqrt{N_0} a$.

For the derivation of $\mathcal{S}^{\perp \perp }$ we use a generating functional
\begin{align}
\mathcal{S}^{\perp\perp}(\mathbf{r}_{ij},t)
=- \frac{1}{4} \frac{d^2}{d \lambda^2}\bra{\psi} &e^{i \lambda (\delta n_i(t) +\delta n_j(t))}\notag \\
&-e^{i \lambda (\delta n_i(t) -\delta n_j(t))}  \ket{\psi}|_{\lambda=0},
\end{align} to bring the expression to a similar form as in $\mathcal{S}^{yy}$. The time evolved operator $\delta n_i(t)$ takes the form $\delta n_i(t)=N_0 \sum_l C'(\mathbf{r}_{il},t) \phi_l - D'(\mathbf{r}_{il},t) \delta n_l$, with $C'(\mathbf{r},t) =  \frac{1}{N}\sum_\mathbf{k}e^{i \mathbf{k}\cdot \mathbf{r}} \sqrt{\frac{V_{\mathbf{k}}}{U_{\mathbf{k}}}}\sin(\omega_{\mathbf{k}}  t) $ and $D'(\mathbf{r},t)= \frac{1}{N} \sum_\mathbf{k} e^{i \mathbf{k}\cdot \mathbf{r}} \cos(\omega_{\mathbf{k}} t)$. We then continue analogously to the computation of $\mathcal{S}^{yy}$ and after performing the derivatives and setting $\lambda=0$ we find 
\begin{align}
\mathcal{S}^{\perp\perp }(\mathbf{r},t)&=\frac{1}{4N}\sum_{\mathbf{k}} \left( 1+\left( \frac{4V_{\mathbf{k}}}{U_{\mathbf{k}}}-1 \right)\sin^2(\omega_\mathbf{k} t) \right)e^{i\mathbf{k}\cdot\mathbf{r}}.
\label{eq:correlationsperp}
\end{align}

\subsection{Comparison with the Linear Spin-Wave Result} 
After performing a Fourier transformation of \cref{eq:correlationsperp} we obtain the result from linear spin wave theory in \cref{eq:corrresultperplsw} (see also \cref{app:comparisonlswandlowenergy}). Since the derivation within low-energy theory does not involve any approximations that break down for short-range interactions this implies, that the linear spin-wave theory provides a good description of the quench dynamics of $\mathcal{S}^{\perp \perp}$ even for short-range interactions. The reason is, that the density type fluctuations are small for all times, independent of the range of the interactions. This result agrees with DMRG-calculations \cite{Frrot2018,Chen2025} and experimental data \cite{Chen2025}, which showed that the early time dynamics of the $S^z$-structure factor is well described by linear spin-wave theory even for short-range interaction. 

Next, we notice that for small $g_{y,1}$ and $g_{y,2}$ the y-correlations in leading order can be approximated by $\mathcal{S}^{yy}(\mathbf{r},t) \propto g_{y,2}(\mathbf{r},t)$ similar to \cref{eq:correlationsperp}. In this case the low-energy result again matches the results from linear spin-wave theory. The assumption that $g_{y,1}(\mathbf{r},t)$ and $g_{y,2}(\mathbf{r},t)$ are small is valid for the weak long range regime, which is also the regime where we expect that the linear spin-wave theory describes the quench dynamics well. However, they are not small in the short-range regime. In the following we thus analyze $\mathcal{S}^{yy}(\mathbf{r},t)$ for short-range interactions in more detail.

\subsection{Phase-Type Correlations for Short-Range Interactions}
\label{sec:shortrangecorr}
Finally, we analyze $\mathcal{S}^{yy}$ for short-range interactions. To this end we compute $g_{y,1}(\mathbf{r},t)$ and  $g_{y,2}(\mathbf{r},t)$.
First, we convert the sums in \cref{eq:correlationsintegral} into an integral and use that for small $\mathbf{k}$ the spectrum $\omega_\mathbf{k}$ depends only on $k=|\mathbf{k}|$ and the coefficient $U_\mathbf{k}$ is approximately constant. Furthermore, for small $k$ the integration is dominated by the term $\propto 1/\omega_\mathbf{k}^2$ and the remaining terms can thus be neglected.
For $\alpha>4$ and $\delta=0$, the resulting integrals can be approximated analytically \cite{tableofintegrals} and one finds $\mathcal{S}^{yy}(r,t)=0$ for $r>2vt$ and for $r<2vt$ 
\begin{align}
\mathcal{S}^{yy}(r,t)&=
  \frac{1}{8 (\Lambda r)^{\frac{\beta_q}{4}}}\left(1-\left(\frac{r}{2vt +\sqrt{(2vt)^2-r^2}}\right)^{\frac{\beta_q}{2}}\right), 
  \label{eq:shortrangecorrana} 
\end{align}
where $v$ is the quasiparticle velocity, $\beta_q=\frac{a^2 U_0^2}{4 \pi v^2}$ and the cutoff $\Lambda$ was introduced to restrict the integration to small $k$-values. Details on the derivation are given in \cref{app:correlationshortrange}.

The result illustrates nicely that correlations in the system cannot propagate faster than $2v$. This can be understood as follows. The correlator measures the probability to find quasiparticles in distance $r$ at time $t$. Initially at $t=0$ the system is in a purely local product state and quasiparticles are pairwise localized at the lattice sites. Each quasiparticle now moves with velocity $v$ in opposite directions and thus the relative velocity between two quasiparticles is $2v$, leading to correlations at maximum distance $2vt$ \cite{Frrot2018}.

\begin{figure}[tb]
  \includegraphics[width= \linewidth]{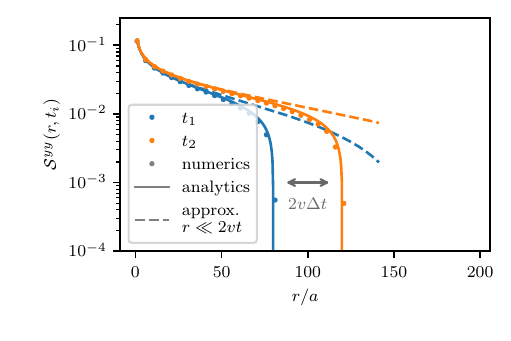}
  \caption{Correlations for nearest neighbor interactions at $v t_1=40a$ and $v t_2=60a$ for $\Lambda=1/a$. The dots are numerically calculated by integrating the exact expressions respecting the angular dependence in $\omega_\mathbf{k}$. The solid blue line corresponds to the analytical result in \cref{eq:shortrangecorrana} and the dashed line corresponds to a small $r$ expansion given by $((r/a)^{-\beta_q/4}-\left(a r/(4 v t)^2\right)^{\beta_q/4})/8$.}
  \label{fig:correlations}
\end{figure}

In figure \cref{fig:correlations} we compare this result with the numerically computed $g_{y,1}(r,t)$ and $g_{y,2}(r,t)$. The approximation fits nicely for $r\ll 2vt$. Small deviations appear for $r\approx 2 v t$. This is due to the fact that in the numerical integration the integral is bounded by $\Lambda=1/a$, while for the analytical result we consider the limit $\Lambda r \rightarrow \infty$. The physical interpretation is that, since each position in the smeared out correlator is averaged over all neighboring positions, the correlations are smoothed and do not sharply drop to 0 at $r=2vt$.

\section{Conclusion}
\label{sec:conclusion}

In summary, we analyzed the quench dynamics of a product state that evolves under a spin-1/2 XY-model with algebraic interactions and a staggered magnetic field. To this end we used linear spin-wave theory and generalized the results of reference \cite{Frrot2018} to arbitrary magnetic field strength. To also cover short-range interactions, we in addition used a low-energy effective theory for phase- and density-fluctuations. For the phase-type correlations this allowed us to derive an analytic expression for the quench dynamics in the short-range regime. Furthermore, our results show that the linear spin-wave result for density-type correlations is valid even in the short-range regime, where some approximations of the linear spin-wave theory are expected to be not valid. This agrees well with recent experiments \cite{Chen2025} and numerical results \cite{Frrot2018,Chen2025} and stresses that the good agreement of the experimental and numerical data with the linear spin-wave result is not restricted to early times, but survives even in the long time limit.

\section*{Acknowledgements}

We would like to thank Antoine Browaeys, Thierry Lahaye, Cheng Chen, Gabriel Emperauger and Guillaume Bornet for insightful discussions. 
This work was funded by the Deutsche Forschungsgemeinschaft (DFG,
German Research Foundation) – project number 554561799 / PF381/24-1.

\newpage
\appendix

\begin{widetext}

\section{Derivation of Quench Dynamics within low-energy Theory}
\label{app:calcmaincorr}

In this section we derive \cref{eq:mainresult}. For convenience, we drop the index $l$ that labels the coarse-grained lattice sites.
First, we use the Baker-Campbell-Hausdorff formula to separate density and phase fluctuations, such that
\begin{align}
\bra{\psi}e^{i N_0 (C\phi-D\delta n)}\ket{\psi}=e^{-i N_0 CD/2}  
\bra{\psi} e^{i N_0 C\phi} e^{-i N_0 D\delta n} \ket{\psi}.
\label{eq:calccorr0}
\end{align}
Here, we used $[\delta n,\phi]=i/N_0$.
The eigenstates $\ket{m}$ of $\delta n$ as well as the eigenstates $\ket{\phi}$ of the operator $\phi$ form a complete basis such that we can write 
\begin{align}
  \mathds{1}&=\int_{-\pi}^\pi \frac{d \phi}{2 \pi} \ket{\phi}\bra{\phi} \notag ,\\
  \mathds{1}&=\sum_{m} \ \ket{m}\bra{m}.
\end{align}
For the states $\ket{m}$ and $\ket{\phi}$ one finds $\braket{\phi|m}=\exp(-i \phi m)$. Using this we can write 
\begin{align}
  \bra{\psi} e^{i N_0 C\phi} e^{-i N_0D\delta n} \ket{\psi}=\sqrt{\frac{2}{\pi N_0}}\int_{-\pi}^\pi \frac{d \phi}{2 \pi}\sum_{m,m'} e^{i(m'-m+N_0 C)\phi}e^{-iDm} e^{-\frac{1}{N_0}m^2} e^{-\frac{1}{N_0}m'^2}.
  \label{eq:derivationcorrelationsmain}
\end{align}
To evaluate the sum over $m'$ in \cref{eq:derivationcorrelationsmain} we use Poisson's summation formula, such that
\begin{align}
\sum_{m'\in \mathbb{Z}} e^{-m'^2/N_0} e^{i m' \phi}
&=\sqrt{N_0 \pi}\sum_{G\in \mathbb{Z}} e^{-N_0(\phi+2 \pi G)^2/4}. 
\end{align}
For large $N_0$ the sum is thus dominated by the $G=0$ term, while all the other terms are exponentially suppressed. 
Inserting the result of the $m'$-summation into \cref{eq:derivationcorrelationsmain} allows us to now perform the integration over $\phi$. Again, since $e^{-N_0 \pi^2/4}$ is small for large $N_0$ we can extend the integration to the entire real axis and thus 
\begin{align}
 \sqrt{N_0 \pi} \int \frac{d \phi}{2 \pi} &e^{-i(m-N_0 C)\phi} e^{-N_0\phi^2/4} = e^{-(m-N_0C)^2/N_0}.
\end{align}
The remaining sum over $m$ can be performed analogously to the sum over $m'$ and the final result for \cref{eq:derivationcorrelationsmain} is given by
\begin{align}
 e^{i\frac{N_0}{2} CD} e^{-\frac{N_0}{2}C^2} e^{-\frac{N_0}{8}D^2}.
\end{align}
The complex prefactor in the first term cancels with the phase we obtained when applying the Baker-Campbell-Hausdorff formula and the final result thus reads 
\begin{align}
\bra{\psi}e^{i N_0 (C\phi-D\delta n)}\ket{\psi}=  e^{-\frac{N_0}{8}\left[4C^2+D^2\right]}.
\label{eq:calccorr02}
\end{align}

\end{widetext}

\section{Comparison of Spin-Wave and Low-Energy Results}
\label{app:comparisonlswandlowenergy}

In this section we compare the results from linear spin-wave theory with the low-energy results. For $2<\alpha<4$ we expect that the results for the correlators we derived with linear spin-wave theory and the result we obtained with the low-energy effective theory approximately match. Here, we verify this and furthermore show that the linear spin-wave results for the perpendicular structure factor provide a qualitatively good description of the quench-dynamics even for short-range interactions. First, we repeat the results from the low-energy theory and derive their Fourier transformation. Afterward, we compare this with the linear spin-wave results. Since we are only interested in the $\omega_+$-mode and the corresponding structure factor, we drop the $+$ here.

We start by repeating the result we obtained for the perpendicular structure factor which is given by
\begin{align}
  \mathcal{S}^{\perp\perp }(\mathbf{r}_{ij},t)&=\frac{1}{4N}\sum_{\mathbf{k}} \left( 1+\left( \frac{4V_{\mathbf{k}}}{U_{\mathbf{k}}}-1 \right)\sin^2(\omega_\mathbf{k} t) \right)e^{i\mathbf{k}\cdot\mathbf{r}_{ij}}.
\end{align}
To bring the spin-spin correlations along $y$ into a similar form, we assume that the exponents in \cref{eq:correlationsb} are small (which is valid for  $2<\alpha<4$) and expand the correlations as 
\begin{align}
\mathcal{S}^{yy}(\mathbf{r}_{ij},t)= \frac{1}{16} g_{y,2}(\mathbf{r}_{ij},t).
\label{eq:correxpansiony}
\end{align}

The structure factors $\mathcal{S}^{\mu \mu}_\mathbf{q}(t)=\frac{1}{N}\sum_{i,j}\mathcal{S}^{\mu \mu}(\mathbf{r}_{ij},t) e^{i \mathbf{q} \mathbf{r}_{ij}}$ then read
\begin{align}
\mathcal{S}_\mathbf{q}^{yy}(t)&\approx\frac{1}{4 }\left(1+\frac{U_\mathbf{q}(U_\mathbf{q}-4 V_\mathbf{q})}{4\omega^2_\mathbf{q}} \sin^2(\omega_\mathbf{q} t)\right)\notag \\
\mathcal{S}_\mathbf{q}^{\perp\perp}(t)&=\frac{ 1}{4} \left(1-\frac{V_\mathbf{q}(U_\mathbf{q}-4 V_\mathbf{q})}{\omega^2_\mathbf{q}} \sin^2(\omega_\mathbf{q} t)\right).
\label{eq:coarsegrainedstructurefactorlowenergy}
\end{align}

For the comparison we repeat the  linear spin-wave results in \cref{eq:corrresultperplsw} 
\begin{align}
\mathcal{S}_\mathbf{q}^{yy}(t)&=\frac{1}{4}\left(1+2 \frac{B_\mathbf{q}(A_\mathbf{q}+B_\mathbf{q})}{\omega^2_\mathbf{q}} \sin^2(\omega_\mathbf{q} t)\right) \notag \\
\mathcal{S}_\mathbf{q}^{\perp\perp}(t)&=\frac{1}{4} \left(1-2  \frac{B_\mathbf{q}(A_\mathbf{q}-B_\mathbf{q})}{\omega^2_\mathbf{q}} \sin^2(\omega_\mathbf{q} t)\right),
\label{eq:coarsegrainedstructurefactorlsw}
\end{align}
where
\begin{align}
  A_\mathbf{k}&=\delta \sin(\chi)+J \gamma_0 \cos^2(\chi)-\frac{J}{2}\gamma_{\mathbf{k}}\notag \\&-\frac{J}{2}\sin^2(\chi)(\gamma_{1,\mathbf{k}}-\gamma_{2,\mathbf{k}})\notag \\
  B_\mathbf{k}&=\frac{J}{2}\gamma_{\mathbf{k}}-\frac{J}{2}\sin^2(\chi)(\gamma_{1,\mathbf{k}}-\gamma_{2,\mathbf{k}}). 
\end{align}
Furthermore, we repeat the coefficent from the low-energy theory 
\begin{align}
U_{\mathbf{k}}&=2 \left(J \gamma_0 \cos^2(\chi) +\delta \sin(\chi)-J \sin^2(\chi)(\gamma_{1,\mathbf{k}}-\gamma_{2,\mathbf{k}})\right)\notag \\ 
V_{\mathbf{k}}&=\frac{1}{2}\left(J \gamma_0 \cos^2(\chi) +\delta \sin(\chi)-J \gamma_{\mathbf{k}}\right)
\end{align}
and identify $U_\mathbf{k}=2 (A_\mathbf{k}+B_\mathbf{k})$, $V_\mathbf{k}=1/2(A_\mathbf{k}-B_\mathbf{k})$ and $U_\mathbf{k}-4 V_\mathbf{k}=4 B_\mathbf{k}$. Inserting this in \cref{eq:coarsegrainedstructurefactorlowenergy} thus gives the same expressions as the linear spin-wave theory in \cref{eq:coarsegrainedstructurefactorlsw}.

\section{Correlations for Short-Range Interactions}
\label{app:correlationshortrange}

Here, we analyze the phase-type correlations for $\alpha>4$. Thus, we have to compute the functions $g_{y,1}(r,t)$ and $g_{y,2}(r,t)$, which are defined in \cref{eq:correlationsintegral}. We start with the computation of $g_{y,1}(r,t)$ and afterward continue with $g_{y,2}(r,t)$.
To obtain an analytic expression for $g_{y,1}(r,t)$ we take the continuum limit and replace the sum over $\mathbf{k}$ by an integration. Furthermore, we use the fact that for small $k$ the Fourier transform of the interaction $\gamma_\mathbf{k}$ and thus also the spectrum $\omega_\mathbf{k}$ depend only on the absolute value of $\mathbf{k}$, which allows us to already perform the angular integration  such that we are left with
\begin{align}
g_{y,1}(r,t) = a^2 &\int_0^\Lambda \frac{dk}{2 \pi} k  (1- J_0(kr))\notag \\
&\times\left[ 4  \cos^2(\omega_k t) +\frac{U_{0}^2}{\omega_{k}^2} \sin^2(\omega_k t)  \right],
\label{eq:hardintegral}
\end{align}
where the cutoff $\Lambda$ ensures the restriction to small $\mathbf{k}$.
The first term in the integral $\propto k (1- J_0(kr)) \cos^2(\omega_k t)$ vanishes for small $k$, and thus we can neglect it when deriving the quench dynamics for large $r$ and $t$.   
For $\delta<\delta_c$, the spectrum for small $k$  takes the form $\omega_k=v k$. Inserting this and using $\sin^2(x)=(1-\cos(2x))/2 $ the integral becomes
\begin{align}
g_{y,1}(r,t) = \beta_q \int_0^\Lambda dk \frac{(1-\cos(2 v k t))}{k} (1- J_0(kr)),
\label{eq:hardintegralinserted}
\end{align}
with the dimensionless constant $\beta_q=\frac{a^2 U_0^2}{4 \pi v^2}$.

The integrand approaches 0 for $k\rightarrow0$ and behaves like $1/k$ for $k \rightarrow \infty$. Thus, one finds the asymptotic behavior 
\begin{align}
g_{y,1}(r,t)  \approx \beta_q \ln(\Lambda r).
\label{eq:hardintegralsola}
\end{align}

Similarly, the derivation of $g_{y,2}(r,t)$ reduces to solving
\begin{align}
g_{y,2}(r,t)=\beta_q \int_0^{\Lambda r} dx \frac{1}{ x} \left(1-\cos\left(\frac{2 v t x}{r}\right)\right)J_0(x),
\label{eq:hardintegralsolhelp}
\end{align}
where we substituted $x=k r$.
For $\Lambda r \rightarrow \infty $ this integral can be solved exactly \cite{tableofintegrals} and one finds
\begin{align}
 g_{y,2}(r,t) &\approx \beta_q \operatorname{arcosh} \left(\frac{2 v t}{r}\right)\Theta\left(\frac{2 v t}{r}-1\right) \notag \\
 &=\beta_q \ln\left(\frac{2 v_s t}{r}+\sqrt{\left(\frac{2 v t}{r}\right)^2-1}\right)\Theta\left(\frac{2 v t}{r}-1\right) . 
\label{eq:hardintegralsolb}
\end{align}
Inserting this into \cref{eq:correlationsb} gives 
\begin{align}
\mathcal{S}^{yy}(r,t)=
  \frac{ (\Lambda r)^{-\frac{\beta_q}{4}}}{8}\left(1-\left(\frac{2vt}{r}+\sqrt{\left(\frac{2vt}{r}\right)^2-1}\right)^{-\frac{\beta_q}{2}}\right).  
\end{align}
The correlations thus decay algebraically for $r\ll 2 v t$ with 
\begin{align}
\mathcal{S}^{yy}(r,t) \approx \frac{1}{8}\left((\Lambda r)^{-\frac{\beta_q}{4}}-\left(\frac{r}{\Lambda (4 v t)^2}\right)^{\frac{\beta_q}{4}}\right).
\label{eq:corrsmallrapprox}
\end{align}
For $\delta=0$ we find $U_0=2 J\gamma_0$ and $v=J \gamma_0 \sqrt{1-\gamma_k/\gamma_0}/k$.

\newpage
\bibliographystyle{bibstyle}

\bibliography{references}

\end{document}